\documentclass[showpacs,draft,preprint,floatfix]{revtex4}
\usepackage{graphicx}

\begin{document}

\preprint{The relative phase gate}
\title{On the bandgap quantum coupler and the harmonic oscillator
interacting with a reservoir: Defining the relative phase gate}
\author{P. C. Garc\'{\i}a Quijas}
\affiliation{Instituto de F\'isica. Universidad de Guanajuato. Loma del Bosque No. 113,
Fracc. Lomas del Campestre, Le\'on, Gto. M\'exico.}
\author{L. M. Ar\'{e}valo Aguilar}
\homepage{http://www.cio.mx}
\email{larevalo@cio.mx}
\affiliation{Centro de Investigaciones en \'Optica, A. C., Loma del Bosque No. 115,
Fracc. Lomas del Campestre, Le\'{o}n, Gto. M\'{e}xico.}
\date{\today }

\begin{abstract}
In order to be able to study dissipation, the interaction between
a single system and their environment was introduced in quantum
mechanics. Master and quantum Langeving equations was derived and,
also, decoherence was studied using this approach. One of the most
used model in this field is a single harmonic oscillator
interacting with a reservoir. In this work we solve analytically
this problem in the resonance case with the evolution operator
method. We use this result to study the conditional dynamics of a
finite system of coupling, a bandgap quantum coupler. We study the
conditional dynamics of the coupler on the computational basis by
choosing a proper interaction time. This conditional dynamics
provides a distinct realization of a quantum phase gate, which we
name the relative phase gate.
\end{abstract}

 \maketitle




\section{Introduction}

One of the most important characteristics of the standard quantum theory is
its restriction to be applied to closed systems, i.e. isolated of
environmental influences, because the Scr\"{o}dinger equation applies only
to closed systems. If we take into account the environment we have an open
system and the total Hamiltonian consists of the Hamiltonian of the open
system, its environment (named reservoir or thermal bath too), and their
interaction:
\begin{equation}
\hat{H}_{tot}=\hat{H}_{sys}+\hat{H}_{env}+\hat{H}_{int}.  \label{bqc1}
\end{equation}%
Particularly, in quantum optics, it is of great practical importance the
study of these systems. For instance, that the quantum system interacting
with its environment becomes entangled with it is considered the profoundly
quantum cause for decoherence. The most simple example of an open system is
a simple harmonic oscillator interacting with an infinity set of harmonics
oscillators. The corresponding hamiltonian, in the rotating-wave
approximation,
\begin{equation}
\hat{H}_{tot}=w\hat{a}^{\dagger }\hat{a}+w\sum_{j=0}^{\infty }\hat{b}%
_{j}^{\dagger }\hat{b}_{j}+\sum_{j=0}^{\infty }g_{j}\left( \hat{a}\hat{b}%
_{j}^{\dag }+\hat{a}^{\dag }\hat{b}_{j}\right) ,  \label{bqc2}
\end{equation}%
describing the resonant interaction between a single harmonic
oscillator and the reservoir oscillators (we have taken $\hbar
=1$). The $g_{j}$ is a real constant describing the linear
coupling. In this case, the Schr\"{o}dinger equation is regarded
as unsolvable and the physical community (particularly the quantum
opticians) have developed many valuable instruments to handle the
problem. For instance, from equation (\ref{bqc2}) the following
master equation in the Lindblad form at temperature $T\neq 0$ is obtained \cite%
{carmichael}:
\begin{equation}
\frac{\partial \hat{\rho}\left( t\right) }{\partial t}=i\omega _{0}\left[
\hat{\rho}\left( t\right) ,\hat{a}^{\dagger }\hat{a}\right] +\frac{\gamma }{2%
}\left( \overline{n}+1\right) \left( 2\hat{a}\hat{\rho}\hat{a}^{\dag }-\hat{a%
}^{\dag }\hat{a}\hat{\rho}-\hat{\rho}\hat{a}^{\dag }\hat{a}\right) +\frac{%
\gamma }{2}\overline{n}\left( 2\hat{a}^{\dag }\hat{\rho}\hat{a}-\hat{a}\hat{a%
}^{\dag }\hat{\rho}-\hat{\rho}\hat{a}\hat{a}^{\dag }\right) ,  \label{bqc3}
\end{equation}%
where $\hat{\rho}\left( t\right) $ is the open system's reduced density
operator. Equation (\ref{bqc3}) has been solved in many ways, see for
example references \cite{carmichael,lu,messina,yo,scully,louisell} and
references therein. Another way to treat this problem is by using the
Heisenberg-Langevin approach \cite{scully,louisell}, where the dynamical
equations are deduced, also, from Hamiltonian (\ref{bqc2}). Both
aproximations, i. e. the master and Langeving equations, focuses in the
single system by tracing out the environmental degrees of freedom \cite%
{carmichael,scully,louisell}. Additionally, the problem of
dissipation is handled by introducing phenomenological decay
constant \cite{dekker,um,onofrio}.

On the other hand, the most straight way to solve the problem is by solving
directly the time dependent Schr\"{o}dinger's equation, i.e.
\begin{equation}
\left\vert \psi \left( t\right) \right\rangle =\hat{U}\left( t\right)
\left\vert \psi \left( 0\right) \right\rangle ,  \label{bqc4}
\end{equation}%
where $\hat{U}\left( t\right) =e^{-\frac{it}{\hbar }\hat{H}_{tot}}=e^{-\frac{%
it}{\hbar }\left( \hat{H}_{sys}+\hat{H}_{env}+\hat{H}_{int}\right) }$ is
called the time evolution operator \cite{ourselve1,ourselve2}. The common
belief of equation (\ref{bqc4}) is that the determination of the solution is
beyond to reach due to the difficult task to factorize $\hat{H}_{tot}$. In
this work, we focus in the analytical solution of this problem by using the
factorization method of the time evolution operator $\hat{U}\left( t\right) $
\cite{ourselve1,ourselve2} (also see reference \cite{nieto}).

We apply this result to the study of conditional dynamics of a finite system
of coupling. The system is an optical coupling device, called the bandgap
quantum coupler, composed of a central waveguide surrounded by a finite
number $N$ of waveguides isolated one from another \cite%
{mogilevtsev,perinajr}. In the limit case when $N\longrightarrow \infty $
the coupler approaches to an open system \cite{mogilevtsev}. We propose a
distinct realization of a quantum phase gate in the coupler, when $N=2,3$,
by identified the optical modes of the waveguides with qubits and selecting
a proper interaction time. We provide this quantum logic gate as an example
of conditional quantum dynamics \cite{nielsen,barenco}.

\section{An harmonic oscillator interacting with the environment}

The coupling between a single harmonic oscillator and the environment is of
great importance in quantum optics. It serves to model an open system and
assist to study the causes of the loss of coherence. The most common way to
determine the open system dynamics is by master equations
\begin{equation}
\frac{\partial \hat{\rho}\left( t\right) }{\partial t}=\hat{\pounds }\hat{%
\rho}\left( t\right) ,  \label{bqc5}
\end{equation}%
where $\hat{\pounds }$ denote the generator of this dissipative, non unitary
dynamics. This equation is of so called Lindblad form, ensuring that general
properties of density operator $\hat{\rho}\left( t\right) $ are preserved
under time evolution. In this section we solve this problem using the
evolution operator method \cite{ourselve1,ourselve2}. This method allow to
find directly the evolution of a single harmonic oscillator coupled with the
environment and avoids having to deal with the equation (\ref{bqc5}).

For a simple harmonic oscillator interacting with the environment the state
system dynamics is given by:
\begin{equation}
\left\vert \Psi \left( t\right) \right\rangle =e^{\epsilon \left( \hat{H}%
_{sys}+\hat{H}_{env}\right) }e^{\epsilon \hat{H}_{int}}\left\vert \Psi
\left( 0\right) \right\rangle ,  \label{bqc6}
\end{equation}%
where $\epsilon \equiv -it$. Evidently the exponential terms are function of
operators wich obey specific commutation rules. In order to obtain $%
\left\vert \Psi \left( t\right) \right\rangle $ we have to separate the
exponential operators. For this purpose, we will use the factorization
method to exponential functions of operators introduced in Refs. \cite%
{ourselve1,ourselve2}. From (\ref{bqc2}) immediately follows that $[\hat{H}%
_{sys}+\hat{H}_{env},\hat{H}_{int}]=0$ and only is necessary the
factorization of interaction terms. Defining $\hat{L}_{+}\equiv \epsilon
\sum_{j=0}^{\infty }g_{j}\hat{a}^{\dagger }\hat{b}_{j}\equiv \hat{L}%
_{-}^{\dagger }$ and $\hat{L}_{3}\equiv \frac{1}{2}\left( \sum_{j=0}^{\infty
}\epsilon ^{2}g_{j}^{2}\hat{a}^{\dagger }\hat{a}-\sum_{i,j=0}^{\infty
}\epsilon ^{2}g_{i}g_{j}\hat{b}_{i}^{\dagger }\hat{b}_{j}\right) $ from the
equation (\ref{bqc2}) we have

\begin{equation}
\left[ \hat{L}_{+},\hat{L}_{-}\right] =2\hat{L}_{3},\left[ \hat{L}_{3},\hat{L%
}_{\pm }\right] =\pm \sum_{j=0}^{\infty }\epsilon ^{2}g_{j}^{2}\hat{L}_{\pm
}.  \label{bqc7}
\end{equation}%
The commutation relations (\ref{bqc7}) obey the standard angular-momentum
commutation relations. In Refs. \cite{ourselve1,ourselve2} the problem has
been solved making use of an unusual symmetrical structure on the
exponential function of operators by repeating $\hat{L}_{+}$ or $\hat{L}_{-}$
on both sides of factorization array and by omitting the commutator $\hat{L}%
_{3}$. This result allow us do not treat with the commutator because the
application of $\exp \left( \sum_{i,j=0}^{\infty }g_{i}g_{j}\hat{b}%
_{i}^{\dagger }\hat{b}_{j}\right) $ is a difficult task due to the
exponential operator contains a double summatory. Therefore, using one of
the two factorization forms found in Refs. \cite{ourselve1,ourselve2} for
the unidimensional harmonic oscillator the equation (\ref{bqc6}) becomes:

\begin{equation}
\left\vert \Psi \left( t\right) \right\rangle =e^{\epsilon w\left( \hat{a}%
^{\dagger }\hat{a}+\sum_{j=0}^{\infty }\hat{b}_{j}^{\dag }\hat{b}_{j}\right)
}e^{\epsilon f\left( t\right) \sum_{j=0}^{\infty }g_{j}\hat{a}^{\dag }\hat{b}%
_{j}}e^{\epsilon h\left( t\right) \sum_{j=0}^{\infty }g_{j}\hat{a}\hat{b}%
_{j}^{\dag }}e^{\epsilon f\left( t\right) \sum_{j=0}^{\infty }g_{j}\hat{a}%
^{\dag }\hat{b}_{j}}\left\vert \Psi \left( 0\right) \right\rangle ,
\label{bqc8}
\end{equation}%
where%
\begin{equation}
f\left( t\right) =\frac{1}{\sqrt{\gamma }}\tan \left( \sqrt{\gamma }%
/2\right) ,h\left( t\right) =\frac{1}{\sqrt{\gamma }}\sin \left( \sqrt{%
\gamma }\right) ,  \label{bqc9}
\end{equation}%
and $\gamma \equiv $ $\sum_{j=0}^{\infty }t^{2}g_{j}^{2}$. There exist
another factorization array for these operator algebras as it is shown in
Refs. \cite{ourselve1,ourselve2} but we will use this array for simplicity
in the application.

At this stage, we have solved formally the time dependent
Schr\"{o}dinger equation for the resonant case of a single
harmonic oscilator interacting with a heat bath. Therefore to
obtain the state system dynamics $\left\vert \Psi \left( t\right)
\right\rangle $ is necessary to choose properly distinct initial
states $\left\vert \Psi \left( 0\right) \right\rangle $. In this
work, we will not treat the problem of state system evolution
under specific conditions on the initial states. Only we show the
mathematical solution of this problem. Instead, we will
concentrate in the conditional quantum dynamics of a finite
coupling system by choosing a specific interaction time and the
initial state in the computation basis $\left\{ 0,1\right\} $.
This result will provide an example of a quantum logic gate for
quantum computation. For an interesting discussion regarding the
applicability of this model and a review of some methods to study
it see the work of Presilla, Onofrio, and Tambini \cite{onofrio}.

\section{The bandgap quantum coupler}

The bandgap quantum coupler is a device that emulate the interaction of a
single harmonic oscillator with a finite set of oscillators. This device is
an optical coupler formed from a central waveguide surrounded by a number
finite $N$ of waveguides isolated one from another, so that it is possible
an interaction with only the central waveguide \cite{mogilevtsev, perinajr}.
We propose this device as a generator of a quantum logical gate by choosing
a suitable interaction time. The Hamiltonian wich describes the system is
given by

\begin{equation}
\hat{H}=w\hat{a}^{\dagger }\hat{a}+w\sum_{j=1}^{N}\hat{b}_{j}^{\dagger }\hat{%
b}_{j}+\sum_{j=1}^{N}g_{j}\left( \hat{b}_{j}\hat{a}^{\dagger }+\hat{a}\hat{b}%
_{j}^{\dagger }\right) ,  \label{bqc10}
\end{equation}%
where $\hat{a}\left( \hat{a}^{\dagger }\right) $ and $\hat{b}_{j}\left( \hat{%
b}_{j}^{\dagger }\right) $ are the corresponding annihilation (creation)
operators of the modes propagating in the central and $jth$ waveguides. The $%
g_{j}=g\left( \forall \text{ }j\right) $ is a real constant describing the
linear coupling. Evidently, this hamiltonian is similar to that of equation (%
\ref{bqc2}), except for the limit of the summatory. Therefore we can obtain
an immediate solution of the corresponding Schr\"{o}dinger equation of the
form:%
\begin{equation}
\left\vert \Psi \left( t\right) \right\rangle =e^{\epsilon w\left( \hat{a}%
^{\dagger }\hat{a}+\sum_{j=1}^{N}\hat{b}_{j}^{\dag }\hat{b}_{j}\right)
}e^{\epsilon f\left( t\right) \sum_{j=1}^{N}g_{j}\hat{a}^{\dag }\hat{b}%
_{j}}e^{\epsilon h\left( t\right) \sum_{j=1}^{N}g_{j}\hat{a}\hat{b}%
_{j}^{\dag }}e^{\epsilon f\left( t\right) \sum_{j=1}^{N}g_{j}\hat{a}^{\dag }%
\hat{b}_{j}}\left\vert \Psi \left( 0\right) \right\rangle ,  \label{bqc11}
\end{equation}%
where $f\left( t\right) $ and $g\left( t\right) $ are the functions of the
equation (\ref{bqc9}). In order to obtain a conditional dynamics wich
implement a quantum logical gate in the coupler, we have to establish a
connection between our system and the language of quantum computation.
Bosonic qubits are defined by states of optical modes. An optical mode is a
physical system whose state space consists of superpositions of the number
states $\left\vert n\right\rangle $, where $n=0,1,...$ gives the number of
photons in the mode \cite{knill,kok}. Then, each optical mode in a waveguide
represents a qubit and the corresponding $\hat{U}\left( t\right) $ generate
the evolutions. Once we have made a suitable connection we will restrict our
study to the case when $N=1$ and the preparation of the initial state $%
\left\vert \Psi \left( 0\right) \right\rangle $ in the two qubits
computational basis $\left\{ \left\vert 0,0\right\rangle ,\left\vert
1,0\right\rangle ,\left\vert 0,1\right\rangle ,\left\vert 1,1\right\rangle
\right\} $.

\subsection{Relative phase gate on two qubits}

Let us show how a two qubits phase gate can be performed using the
interaction (\ref{bqc11}) when $N=1$. This gate is a distinct
realization of the known quantum phase gates \cite{nielsen}.
Initially we consider the case $N=1$ , i.e. two waveguides, and
prepare the initial state in the two qubits computational basis. We
will use the notation $\left\vert n\right\rangle _{a} $ for the
central waveguide and $\left\vert n\right\rangle _{b_{1}}$for first
outer waveguide ($n=0,1$). For the particular interaction time
$t=2\pi /cg$ and $w=cg/2$ the sequence of final states of the
interacting modes is given by
\begin{eqnarray}
\left\vert 0\right\rangle _{a}\left\vert 0\right\rangle _{b_{1}}
&\longrightarrow &\left\vert 0\right\rangle _{a}\left\vert 0\right\rangle
_{b_{1}},  \nonumber \\
\left\vert 1\right\rangle _{a}\left\vert 1\right\rangle _{b_{1}}
&\longrightarrow &\left\vert 1\right\rangle _{a}\left\vert 1\right\rangle
_{b_{1}},  \nonumber \\
\left\vert 1\right\rangle _{a}\left\vert 0\right\rangle _{b_{1}}
&\longrightarrow &e^{i\pi }\left\vert 1\right\rangle _{a}\left\vert
0\right\rangle _{b_{1}},  \nonumber \\
\left\vert 0\right\rangle _{a}\left\vert 1\right\rangle _{b_{1}}
&\longrightarrow &e^{i\pi }\left\vert 0\right\rangle _{a}\left\vert
1\right\rangle _{b_{1}}.  \label{bqc12}
\end{eqnarray}%
The effect of this interaction is to change the phase if both qubits
are in different states. Otherwise the qubits are left unchanged.
From this result, we define the following conditional phase gate
through the mathematical equation:
\begin{equation}
\hat{U}_{Phase}^{relative}\left\vert j_{1}\right\rangle \left\vert
j_{2}\right\rangle =e^{i\pi \left( j_{1}-j_{2}\right) }\left\vert
j_{1}\right\rangle \left\vert j_{2}\right\rangle ,  \label{bqc13}
\end{equation}%
for $j_{1},j_{2}=0,1$. We will call $\hat{U}_{Phase}^{relative}$ as
the \textit{relative phase} gate \cite{nielsen,fiurasek}. In summary
we have made a new version of a quantum phase gate operating on the
$2^{2}$ quantum computational states. Below we give the conditional
definition of this phase gate.

Before to analyze this new phase gate, let us ask: What kind of
evolution is expected when a two-qubit phase gate acts on two
unknown qubits?. That is, the question is:
$\hat{U}_{Phase}^{two-qubit}\left(\alpha|0\rangle_1+\beta|1\rangle_1\right)\left(\gamma|0\rangle_2+\delta|1\rangle_2\right)
\quad\rightarrow\quad?\quad ?$. The definition of one-qubit phase
gate can serve as a guide to  answer this question, it is defined by
its action on the one qubit computational basis states: if the qubit
is in the state $|1\rangle$, then it applies a phase, that is
$\hat{U}_{Phase}^{one-qubit}|1\rangle=e^{i\phi}|1\rangle$; and it
does nothing when the qubit is in the state $|0\rangle$, that is
$\hat{U}_{Phase}^{one-qubit}|0\rangle=|0\rangle$ \cite{nielsen}.
Therefore, when the one-qubit phase gate acts on an unknown qubit,
it produces the evolution:
$\hat{U}_{Phase}^{one-qubit}\left(\alpha|0\rangle+\beta|1\rangle\right)=\left(\alpha|0\rangle-\beta|1\rangle\right)$.
Then, from this definition, we can infer that a two-qubit phase gate
must induce the following conditional evolution:
\begin{eqnarray}
\hat{U}_{Phase}^{two-qubit}\left(\alpha|0\rangle_1+\beta|1\rangle_1\right)\left(\gamma|0\rangle_2+\delta|1\rangle_2\right)
=\left(\alpha|0\rangle_1-\beta|1\rangle_1\right)\left(\gamma|0\rangle_2-\delta|1\rangle_2\right)
\end{eqnarray}

Now, let us review some properties of others two-qubit phase gates
defined in the literature \cite{nielsen}. As far as we know, there
are two two-qubits phase gates defined. The first case, which is
called the control C phase gate, or Cirac-Zoller phase gate
\cite{berman}, $\hat{U}_{Phase}^{C}$ , is defined as: \textit{If
the control qubit is set to $|1\rangle$, then apply the one-qubit
phase gate to the target qubit. Otherwise, if the control qubit is
set to $|0\rangle$, then left the target qubit unchanged.} This
definition produces the following evolution:
$\hat{U}_{Phase}^{C}|0\rangle_1|0\rangle_2=|0\rangle_1|0\rangle_2$,
$\hat{U}_{Phase}^{C}|0\rangle_1|1\rangle_2=|0\rangle_1|1\rangle_2$,
$\hat{U}_{Phase}^{C}|1\rangle_1|0\rangle_2=|1\rangle_1\hat{U}_{Phase}^{one-qubit}|0\rangle_2=|1\rangle_1|0\rangle_2$,
$\hat{U}_{Phase}^{C}|1\rangle_1|1\rangle_2=|1\rangle_1\hat{U}_{Phase}^{one-qubit}|1\rangle_2=|1\rangle_1\left(-|1\rangle_2
\right)=-|1\rangle_1|1\rangle_2$. In short:
\begin{equation}
\hat{U}_{Phase}^{C}|m\rangle_1|n\rangle_2=|m\rangle_1\left(e^{imn\pi}|n\rangle_2\right),\label{CZ-phase}\end{equation}
where, $m,n=0,1$.

In the second case, which we call the control phase shift gate
$\hat{U}_{Phase}^{shift}$, the definition given on page $180$ of
reference \cite{nielsen} is as follows: \textit{If the control qubit
is set to $|0\rangle$, then the target qubit is left alone.
Otherwise, if the control qubit is set to $|1\rangle$, then apply a
phase shift to the target qubit}. This definition produces the
following evolution
$\hat{U}_{Phase}^{shift}|0\rangle_1|0\rangle_2=|0\rangle_1|0\rangle_2$,
$\hat{U}_{Phase}^{shift}|0\rangle_1|1\rangle_2=|0\rangle_1|1\rangle_2$,
$\hat{U}_{Phase}^{shift}|1\rangle_1|0\rangle_2=|1\rangle_1\left(-|0\rangle_2\right)=-|1\rangle_1|0\rangle_2$,
$\hat{U}_{Phase}^{shift}|1\rangle_1|1\rangle_2=|1\rangle_1\left(-|1\rangle_2\right)=-|1\rangle_1|1\rangle_2$.
In short:
\begin{equation}
\hat{U}_{Phase}^{shift}|m\rangle_1|n\rangle_2=|m\rangle_1\left(e^{im\pi}|n\rangle_2\right),\label{c-phase}\end{equation}
where, $m,n=0,1$. Note that the phase induced by this gate depends
exclusively on the value $m$ of the control qubit. Other
characteristic of this phase gate is that, contrary to the one-qubit
phase gate $U_{Phase}^{one-qubit}$, it applies a phase to the base
state $|0\rangle$.

On the other hand, the control phase shift gate
$\hat{U}_{Phase}^{shift}$ suggest, seemingly, a change on the target
qubit. However, when this phase gate is applied to two unknown
qubits it produces the following change:
\begin{equation}
\hat{U}_{Phase}^{shift}\left(\alpha|0\rangle_1+\beta|1\rangle_1\right)\left(\gamma|0\rangle_2+\delta|1\rangle_2\right)
=\left(\alpha|0\rangle_1-\beta|1\rangle_1\right)\left(\gamma|0\rangle_2+\delta|1\rangle_2\right).\label{apply-c-phase}
\end{equation}
Equation (\ref{apply-c-phase}) implies
$\hat{U}_{Phase}^{shift}|0\rangle_1\left(\gamma|0\rangle_2+\delta|1\rangle_2\right)
=|0\rangle_1\left(\gamma|0\rangle_2+\delta|1\rangle_2\right)$, and
$\hat{U}_{Phase}^{shift}|1\rangle_1\left(\gamma|0\rangle_2+\delta|1\rangle_2\right)
=-|1\rangle_1\left(\gamma|0\rangle_2+\delta|1\rangle_2\right)$. It
seems as if this gate produces a phase change only in the control
qubit.

Also, when we apply the control C phase gate on two unknown qubits
we obtain the following entangled state:
\begin{equation}
\hat{U}_{Phase}^{C}\left(\alpha|0\rangle_1+\beta|1\rangle_1\right)\left(\gamma|0\rangle_2+\delta|1\rangle_2\right)
=\alpha|0\rangle_1\left(\gamma|0\rangle_2+\delta|1\rangle_2\right)+
\beta|1\rangle_1\left(\gamma|0\rangle_2-\delta|1\rangle_2\right).\label{ApplyCZ-phase}
\end{equation}
Equation (\ref{ApplyCZ-phase}) implies that
$\hat{U}_{Phase}^{C}|0\rangle_1\left(\gamma|0\rangle_2+\delta|1\rangle_2\right)
=|0\rangle_1\left(\gamma|0\rangle_2+\delta|1\rangle_2\right)$, and
that
$\hat{U}_{Phase}^{C}|1\rangle_1\left(\gamma|0\rangle_2+\delta|1\rangle_2\right)
=|1\rangle_1\left(\gamma|0\rangle_2-\delta|1\rangle_2\right)$. That
is, the control C phase gate induces a phase change only when the
first qubit is in the state $|1\rangle_1$ and the second is in an
unknown state. Because its ability to produce entangled sates, the
control C phase gate (and its three qubit generalization) is the
most studied
\cite{turchette,guozou,zou2,zubairy,kiesel,charron,azuma,ekert1,jones,vager,yang,ekert2}
of the phase gate's family.

On the other hand, the relative phase gate,
$\hat{U}_{Phase}^{relative}$, can be stated in conditional sentence
of ''If-Then'' form as follows:
\begin{quote}
\textsf{\textsc{Definition 1.} If the states of a two parties
system are not equal, then apply the one-qubit phase gate,
$U_{Phase}^{one-qubit}$, on each qubit. Otherwise, left them
unchanged.}\end{quote} Where, by equal states we means:
$|0\rangle_1|0\rangle_2$ or $|1\rangle_1|1\rangle_2$. It is
worthwhile to note that the conditional property of this phase
gate is given by the whole state of the system, which represents a
conditional evolution that depend on the overall state of the
whole system and induces a conditional evolution on each
subsystem, see reference \cite{quijas-are}. \textsc{Definition 1}
produces the following conditional evolution on the computational
basis of two qubits \Big(here we let the one-qubit phase gate to
apply an arbitrary phase $\theta$, that is
$\hat{U}_{Phase}^{one-qubit}|1\rangle= e^{i\theta}|1\rangle$\Big):
$\hat{U}_{Phase}^{relative}|0\rangle_1|0\rangle_2=|0\rangle_1|0\rangle_2,$
$\hat{U}_{Phase}^{relative}|1\rangle_1|1\rangle_2=|1\rangle_1|1\rangle_2,$
$\hat{U}_{Phase}^{relative}|0\rangle_1|1\rangle_2=\hat{U}_{Phase}^{one-qubit}|0\rangle_1
\hat{U}_{Phase}^{one-qubit}|1\rangle_2=
e^{i\theta}|0\rangle_1|1\rangle_2$,
$\hat{U}_{Phase}^{relative}|1\rangle_1|0\rangle_2=\hat{U}_{Phase}^{one-qubit}|1\rangle_1\hat{U}_{Phase}^{one-qubit}|0\rangle_2=
e^{i\theta}|1\rangle_1|0\rangle_2.$

 Now, when we apply the
relative phase gate $\hat{U}_{Phase}^{relative}$ to two unknown
qubits we obtain the following entangled state:
\begin{equation}
\hat{U}_{Phase}^{relative}\left(\alpha|0\rangle_1+\beta|1\rangle_1\right)\left(\gamma|0\rangle_2+\delta|1\rangle_2\right)=
\alpha|0\rangle_1\left(\gamma|0\rangle_2+\delta
e^{i\theta}|1\rangle_2\right)+\beta e^{i\theta}|1\rangle_1
\left(\gamma|0\rangle_2+\delta e^{-i\theta}|1\rangle_2\right)
\end{equation}
For the particular case $\theta=\pi$ we obtains:
\begin{equation}
\hat{U}_{Phase}^{relative}\left(\alpha|0\rangle_1+\beta|1\rangle_1\right)\left(\gamma|0\rangle_2+\delta|1\rangle_2\right)
=\left(\alpha|0\rangle_1-\beta|1\rangle_1\right)\left(\gamma|0\rangle_2-\delta|1\rangle_2\right)\label{teta1}.
\end{equation}
Furthermore, equation (\ref{teta1}) implies that
$\hat{U}_{Phase}^{relative}|0\rangle_1\left(\gamma|0\rangle_2+\delta|1\rangle_2\right)=
|0\rangle_1\left(\gamma|0\rangle_2-\delta|1\rangle_2\right) $ and
$\hat{U}_{Phase}^{relative}|1\rangle_1\left(\gamma|0\rangle_2+\delta|1\rangle_2\right)=-
|1\rangle_1\left(\gamma|0\rangle_2-\delta|1\rangle_2\right) $. Form this result, we can say that the $\hat{U}%
_{Phase}^{relative}$ gate induce a phase change when acts on two
unknown qubits.

It is interesting to note that the relative phase gate in equation
(\ref{teta1}) can be implemented using the control phase shift and
SWAP gates, i. e.
$\hat{U}_{Phase}^{relative}=\hat{U}_{Phase}^{shift}\hat{U}_{SWAP}\hat{U}_{Phase}^{shift}\hat{U}_{SWAP}$.

\subsection{Three relative phase gate}

In the previous section we have proposed a distinct phase gate $\hat{U}%
_{Phase}^{relative}$ in terms of a conditional dynamics on the
global state system. Moreover, a serie of results concerning the
classification of two qubits phase gates from conditional dynamics
were presented. In the following we will generalize the result of
the $\hat{U}_{Phase}^{relative}$ gate to the case of a three qubits
gate. Consider now the case $N=2$, i.e.
three waveguides, in equation (\ref{bqc10}) and the initial state $%
\left\vert \Psi \left( 0\right) \right\rangle $ prepared in the three qubits
computational basis $\left\{ \left\vert 0,0,0\right\rangle ,\left\vert
1,0,0\right\rangle ,\left\vert 0,0,1\right\rangle ,\left\vert
0,1,0\right\rangle ,\left\vert 1,1,0\right\rangle ,\left\vert
1,0,1\right\rangle ,\left\vert 0,1,1\right\rangle ,\left\vert
1,1,1\right\rangle \right\} $. We will use the notation $\left\vert
n\right\rangle _{a}$ for the central waveguide and $\left\vert
n\right\rangle _{b_{1}}\left\vert n\right\rangle _{b_{2}}$ for the first and
second outer waveguides ($n=0,1$). For the specific interaction time $t=2\pi
/cg$ and $w=cg/2$ the sequence of final states of the interacting modes is
given as follows:
\begin{eqnarray}
\left\vert 0\right\rangle _{a}\left\vert 0\right\rangle _{b_{1}}\left\vert
0\right\rangle _{b_{2}} &\longrightarrow &\left\vert 0\right\rangle
_{a}\left\vert 0\right\rangle _{b_{1}}\left\vert 0\right\rangle _{b_{2}},
\nonumber \\
\left\vert 0\right\rangle _{a}\left\vert 1\right\rangle _{b_{1}}\left\vert
1\right\rangle _{b_{2}} &\longrightarrow &\left\vert 0\right\rangle
_{a}\left\vert 1\right\rangle _{b_{1}}\left\vert 1\right\rangle _{b_{2}},
\nonumber \\
\left\vert 1\right\rangle _{a}\left\vert 0\right\rangle _{b_{1}}\left\vert
1\right\rangle _{b_{2}} &\longrightarrow &\left\vert 1\right\rangle
_{a}\left\vert 0\right\rangle _{b_{1}}\left\vert 1\right\rangle _{b_{2}},
\nonumber \\
\left\vert 1\right\rangle _{a}\left\vert 1\right\rangle _{b_{1}}\left\vert
0\right\rangle _{b_{2}} &\longrightarrow &\left\vert 1\right\rangle
_{a}\left\vert 1\right\rangle _{b_{1}}\left\vert 0\right\rangle _{b_{2}},
\nonumber \\
\left\vert 1\right\rangle _{a}\left\vert 0\right\rangle _{b_{1}}\left\vert
0\right\rangle _{b_{2}} &\longrightarrow &e^{i\pi }\left\vert 1\right\rangle
_{a}\left\vert 0\right\rangle _{b_{1}}\left\vert 0\right\rangle _{b_{2}},
\nonumber \\
\left\vert 0\right\rangle _{a}\left\vert 1\right\rangle _{b_{1}}\left\vert
0\right\rangle _{b_{2}} &\longrightarrow &e^{i\pi }\left\vert 0\right\rangle
_{a}\left\vert 1\right\rangle _{b_{1}}\left\vert 0\right\rangle _{b_{2}},
\nonumber \\
\left\vert 0\right\rangle _{a}\left\vert 0\right\rangle _{b_{1}}\left\vert
1\right\rangle _{b_{2}} &\longrightarrow &e^{i\pi }\left\vert 0\right\rangle
_{a}\left\vert 0\right\rangle _{b_{1}}\left\vert 1\right\rangle _{b_{2}},
\nonumber \\
\left\vert 1\right\rangle _{a}\left\vert 1\right\rangle _{b_{1}}\left\vert
1\right\rangle _{b_{2}} &\longrightarrow &e^{i\pi }\left\vert 1\right\rangle
_{a}\left\vert 1\right\rangle _{b_{1}}\left\vert 1\right\rangle _{b_{2}}.
\label{bqc14}
\end{eqnarray}%
The effect of the interaction is to change the sign of the global
state if one or all qubits are in the first excitated state
$\left\vert 1\right\rangle $. Otherwise the qubits are left
unchanged. From this result, we represent the conditional three
qubits phase gate through the mathematical equation:
\begin{equation}
\hat{U}_{Phase}^{relative}\left\vert j_{1}\right\rangle \left\vert
j_{2}\right\rangle \left\vert j_{3}\right\rangle =e^{i\pi \left(
j_{1}-j_{2}-j_{3}\right) }\left\vert j_{1}\right\rangle \left\vert
j_{2}\right\rangle \left\vert j_{3}\right\rangle ,  \label{bqc15}
\end{equation}%
for all $j_{1},j_{2},j_{3}=0,1$. The three qubits relative phase gate, $\hat{%
U}_{Phase}^{relative}$, can be stated in the form of a conditional sentence
\textquotedblright If-Then\textquotedblright\ as follows:

\begin{quote}
\textsf{\textsc{Definition 2.} If the states of a three parties
system are not in the basic sate, i. e.
$|0\rangle_1|0\rangle_2|0\rangle_3$ or an equivalent state, then
apply the one-qubit phase gate, $U_{Phase}^{one-qubit}$, on each
qubit. Otherwise, left them unchange}
\end{quote}

In this sense, we have made a conditional property of the three qubits $\hat{%
U}_{Phase}^{relative}$ on the whole state of the system, i.e. a condition on
the global sate of the system through the values of $j_{1}$, $j_{2}$ and $%
j_{3}$.

Now, when we apply the three qubits relative phase gate $\hat{U}%
_{Phase}^{relative}$ to three unknown qubits we obtain the following result:
\begin{eqnarray}
\hat{U}_{Phase}^{relative}\left( \alpha |0\rangle _{1}+\beta
|1\rangle _{1}\right) \left( \gamma |0\rangle _{2}+\delta
|1\rangle _{2}\right) \left( \eta |0\rangle _{3}+\zeta |1\rangle
_{3}\right)\nonumber \\  = \left( \alpha |0\rangle _{1}-\beta
|1\rangle _{1}\right) \left( \gamma |0\rangle _{2}-\delta
|1\rangle _{2}\right) \left( \eta |0\rangle _{3}-\zeta |1\rangle
_{3}\right)\label{applyr3-phase},
\end{eqnarray}
according to equation (\ref{bqc15}). The equation
(\ref{applyr3-phase}) implies that $|0\rangle _{1}\left( \gamma
|0\rangle _{2}+\delta |1\rangle _{2}\right) \left( \eta |0\rangle
_{3}+\zeta |1\rangle _{3}\right) \rightarrow |0\rangle _{1}\left(
\gamma |0\rangle _{2}-\delta |1\rangle
_{2}\right) \left( \eta |0\rangle _{3}-\zeta |1\rangle _{3}\right) $ and $%
|1\rangle _{1}\left( \gamma |0\rangle _{2}+\delta |1\rangle _{2}\right)
\left( \eta |0\rangle _{3}+\zeta |1\rangle _{3}\right) \rightarrow e^{i\pi
}|1\rangle _{1}\left( \gamma |0\rangle _{2}-\delta |1\rangle _{2}\right)
\left( \eta |0\rangle _{3}-\zeta |1\rangle _{3}\right) $. From this result,
we can say that the three qubits $\hat{U}_{Phase}^{relative}$ gate induce a
phase change when acts on three unknown qubits, as it was expected from the
extension of two qubits $\hat{U}_{Phase}^{relative}$ gate.

\section{Conclusions}

In this work we have addressed the problem of interaction between an
harmonic oscillator and a reservoir. We have found an analytical
solution to the Schr\"{o}dinger equation in the resonant case. Also,
we have presented an explicit example of a finite coupling device
wich simulate the dynamics of an open system in the limit
$N\rightarrow \infty $. The coupler is an optical device of
waveguides concentrated around a central waveguide. Such a
configuration allow us\ to find particular applications by choosing
properly the interaction time. In the cases $N=2,3$ we have found a
conditional dynamics wich provides an example of a quantum logic
gate. This gate adhere a phase on the global state according to a
conditional logic operation. This gate let us to find important
differences with the control C phase gate and the control shift
phase gate. We believe that this result may be useful to establish a
classification of two qubits logic gates in two classes. Gates wich
act on a single qubit and gates wich act on two qubits depending on
a conditional operation \cite{quijas-are}.

\acknowledgments One of us, P. C. Garcia Quijas thanks the support
by Consejo Nacional de Ciencia y Tecnolog\'{\i}a (CONACYT). L. M.
Ar\'{e}valo Aguilar thanks to Sistema Nacional de Investigadores
(SNI) of Mexico.


\end{document}